\documentclass[11pt,twocolumn,twoside]{IEEEtran}
\usepackage{amsmath}
\usepackage[margin=0.75in,headheight=0.45in]{geometry}
\usepackage[pdftex]{epsfig}
\usepackage{amsfonts}
\usepackage{amssymb}
\usepackage{fancyhdr}
\usepackage{url}
\usepackage{xcolor}
\include{graphicsx}

\begin{document}
\title{\vspace{0.2in}\sc Towards Unsupervised Segmentation of Extreme Weather Events}
\author{Adam Rupe$^{1}$\thanks{Corresponding author: A. Rupe, atrupe@ucdavis.edu $^1$Complexity Sciences Center and Department of Physics, University of California Davis $^2$NERSC, Lawrence Berkeley National Laboratory $^3$ Intel Corporation}, Karthik Kashinath$^{2}$, Nalini Kumar$^{3}$, Victor Lee$^{3}$, Prabhat $^{2}$, James P. Crutchfield $^{1}$}

\maketitle
\thispagestyle{fancy}
\begin{abstract}
Extreme weather is one of the main mechanisms through which climate change will directly impact human society. Coping with such change as a global community requires markedly improved understanding of how global warming drives extreme weather events. While alternative climate scenarios can be simulated using sophisticated models, identifying extreme weather events in these simulations requires automation due to the vast amounts of complex high-dimensional data produced. Atmospheric dynamics, and hydrodynamic flows more generally, are highly structured and largely organize around a lower dimensional skeleton of coherent structures. Indeed, extreme weather events are a special case of more general hydrodynamic coherent structures. We present a scalable physics-based representation learning method that decomposes spatiotemporal systems into their structurally relevant components, which are captured by latent variables known as \emph{local causal states}. For complex fluid flows we show our method is capable of capturing known coherent structures, and with promising segmentation results on CAM5.1 water vapor data we outline the path to extreme weather identification from unlabeled climate model simulation data.
\end{abstract}

\section{Introduction}

Life across the globe has survived and thrived by adapting to its local weather, including extreme events such as strong winds and floods from cyclones, drought and heat waves from blocking events and large-scale atmospheric oscillations, and critically-needed precipitation from atmospheric rivers. Driven by an ever-warming climate, extreme weather events are changing in frequency and intensity at an unprecedented pace~\cite{Eman87a,Webs05a}. We need to understand these events and their driving mechanisms to enable communities to continue to adapt and thrive.

High-resolution, high-fidelity global climate models are an indispensable tool for investigating climate change. A multitude of climate change scenarios are now being simulated, each producing 100s of TBs of data. Currently, climate change is assessed in these simulations using summary statistics such as mean global sea surface temperature. This is inadequate for answering detailed questions about the effects of climate change on extreme weather events. Due to the sheer size and complexity of these simulated data sets, it is essential to develop robust and automated methods that can provide the deeper insights we seek.


Recently, supervised Deep Learning (DL) techniques have been applied to address this problem~\cite{mudi17a, kurt18a, jian18a, cohe19d}. Further progress however has been stymied by two daunting challenges: reliance on labeled training data and interpretability of trained models. The DL models used in the above studies are trained using the automated heuristics of TECA~\cite{Prab12a} for proximate labels. This is necessary because, simply put, there currently is no ground truth for pixel-level identification of extreme weather events~\cite{Shie18a}.
While the results in \cite{mudi17a} show that DL can improve upon TECA, the results of \cite{ cohe19d} reach accuracy rates over 97\% and thus essentially just reproduce the output of TECA. The supervised learning paradigm of optimizing 
objective metrics (e.g. training and generalization error)
breaks down here \cite{Fagh14a}; TECA is not ground truth and we do not know how to train a DL model to disagree with TECA in just the right way to get closer to ``ground truth". 

To avoid this issue, a campaign is currently underway to generate expert-labeled training data~\cite{climatenet}. Supervised DL models trained on this data will automate expert-level curation of large climate data sets for extreme weather detection. 
In this case there too will be challenges. Though an improvement over automated heuristics, expert-labeled data is still not an objective ground truth. Further, while human experts can debate the subtleties of physical characteristics of extreme weather events, the interpretability problem~\cite{Olah18a} prevents us from probing a trained DL model to determine exactly how and why it identifies (or misidentifies) specific events. 



To circumvent these challenges of DL-based approaches, here we take an alternative physics-based unsupervised approach, complementary to DL.

Whereas DL takes inspiration from the human visual processing system to identify patterns in images without consideration for what constitutes a ``pattern", our method builds on a theory that seeks to understand the physical nature of pattern without consideration for the visual system that can readily identify such patterns. 
When viewing a video of a complex fluid flow we do not track the evolution of each individual pixel. Our vision instead focuses on relatively few collective features, generally referred to as \emph{coherent structures}~\cite{Holm12a, Hall15a}, that the flow organizes around. Beyond fluid flows, coherent structures in spatiotemporal systems can similarly be understood as key organizing features that heavily dictate the dynamics of the full system, and thus provide a natural dimensionality reduction. 
Understanding the lower-dimensional coherent structures gets us most of the way to understanding and predicting the full higher-dimensional system, and, as with extreme weather, the coherent structures are often the features of interest. 

Our approach seeks to understand the physical nature of coherent structures so that we can discover and identify them in spatiotemporal systems.
It is difficult however, if not impossible, to give an actionable definition of coherent structures as the solution of a general mathematical coherence principle derived from equations of motion. 
Identifying and predicting complex emergent behaviors starting from fundamental laws is typically infeasible \cite{Ande72a}. 
For example, despite knowing the equations of hydrodynamics and thermodynamics, which critically govern the dynamics of hurricanes, many aspects of how hurricanes form and evolve are still poorly understood \cite{Eman03a}.

As a response, research on complex, nonlinear systems shifted to focus directly on system behaviors rather than governing equations.The resulting \emph{behavior-driven theories} (e.g. \cite{Will15a, Rung15a, Rubi18a, Zeni19a}), which lie at the interface of physics and machine learning, provide a new means of scientific discovery directly from data. Our approach to unsupervised extreme weather event detection is through a behavior-driven theory of coherent structures in spatiotemporal systems. Below we give some basics of the theory then demonstrate its utility by identifying known coherent structures in 2D turbulence simulation data and observational data of Jupiter's clouds from the NASA Cassini spacecraft. Finally, we show promising results on CAM5.1 water vapor data and outline the path to extreme weather event segmentation masks. 

\section{Method: Local Causal States}


Our behavior-driven theory of coherent structures builds on a more general theory of pattern and structure in natural systems.
A quantitative theory of structure need be probabilistic, capturing structure in ensembles of behavior, and algebraic, generalizing from the group-theoretic formalism of exact symmetry to the semi-group algebra of finite-state machines.
The mathematical representation of the structure of a system's dynamical behavior is given by a minimal, optimally predictive, stochastic model~\cite{Wolf84a,Gras86,Shal98a}.
For a model to optimally predict with minimal resources that model must capture pattern and structure present in the system's behaviors.

Computational mechanics~\cite{Crut12a} makes this idea operational through the \emph{causal equivalence relation};

\scalebox{0.88}{%
  \hspace*{-5mm}
  $
    \mathrm{past}_i \sim_\epsilon \mathrm{past}_j \iff \Pr(\mathrm{Future} | \mathrm{past}_i) = \Pr(\mathrm{Future} | \mathrm{past}_j)
    ~.
  $
}
The equivalence classes over pasts induced by the causal equivalence relation are known as the \emph{causal states} of the system; they are the unique minimal sufficient statistic of the past for optimally predicting the future. 

For spatiotemporal systems, \emph{lightcones} are used as local notions of past and futures. Two past lightcones $\ell^-_i$ and $\ell^-_j$ are causally equivalent if they have the same conditional distribution over future lightcones;
$$
\ell^-_i \sim_\epsilon \ell^-_j \iff \Pr(\mathrm{L}^+ | \ell^-_i) = \Pr(\mathrm{L}^+ | \ell^-_j)
~.
$$
The resulting equivalence classes are called \emph{local causal states} \cite{Shal03a}. They are the unique minimal sufficient statistic of past lightcones for optimal prediction of future lightcones.
The $\epsilon$-function, which generates the causal equivalence classes, maps from past lightcones to local causal states; $\epsilon: \ell^- \mapsto \xi$. 
Segmentation is achieved by mapping a spacetime field $X$ to its associated local causal state field $S = \epsilon(X)$: every feature $x = X(\vec{r},t)$ is mapped to its classification label (local causal state) via its past lightcone $\xi = S(\vec{r}, t) = \epsilon\bigr(\ell^-(\vec{r}, t)\bigl)$.
Crucially, this ensures the global latent variable field $S$ maintains the same geometry of $X$ such that $S(\vec{r}, t)$ is the local latent variable corresponding to the local observable $X(\vec{r}, t)$. 

For real-valued systems, such as the fluid flows considered here, local causal state reconstruction requires a discretization to empirically estimate $\Pr(\mathrm{L}^+ | \ell^-)$~\cite{Goer12a}. We use K-Means to cluster over lightcones with the lightcone distance metric
$$
\mathrm{D}_{\mathrm{lc}}(\mathbf{a}, \mathbf{b}) \equiv \sqrt{(a_1 - b_1)^2 + \ldots + \mathrm{e}^{-\tau d(n)}(a_n - b_n)^2}
~,
$$
where $\mathbf{a}$ and $\mathbf{b}$ are flattened lightcone vectors, $d(n)$ is the temporal depth of the lightcone vector at index $n$, and $\tau$ is the temporal decay rate ($1/\tau$ can be thought of as a coherence time). 

\section{Results: Structural Segmentation}

Because the local causal state latent variables are designed to capture structure in spatiotemporal systems, we call this level of segmentation a \emph{structural segmentation}. 
That is, the classification assignments are labels of unique local causal states. This is an intermediate step for coherent structure segmentation (e.g. classification assignments of `cylcone', `atmospheric river', `background', etc.), for which an additional layer of analysis is needed on top of the local causal states~\cite{Rupe18b}. 
From comparison with established Lagrangian Coherent Structure results we will now show that physically meaningful coherent structures are captured by our structural segmentation, and then we outline how extreme weather events may be extracted from structural segmentation of global climate data. 


Building on the realization that relatively low-dimensional chaotic attractors underlies turbulent fluid flows~\cite{Ruel71a, Bran83}, the Lagrangian Coherent Structure (LCS) approach is grounded in nonlinear dynamical systems theory and seeks to describe the most repelling, attracting, and shearing material surfaces that form the skeletons of Lagrangian particle dynamics \cite{Hall15a}. LCS are conjectured to capture localized, emergent structures that organize the large-scale flow. We directly compare our results with the geodesic and LAVD approaches (described below) on the 2D turbulence data set from \cite{Hadj17a} and the Jupiter data set from \cite{Hadj17a} and \cite{Hadj16a}.

There are three classes of flow structures in the LCS framework;
elliptic LCS are rotating vortex-like structures, parabolic LCS are generalized Lagrangian jet-cores, and hyperbolic LCS are tendril-like stable-unstable manifolds in the flow. 
The geodesic approach \cite{Hall15a, Hadj16a} is the state-of-the-art method designed to capture all three classes of LCS and has a nice interpretation for the structures it captures in terms of characteristic deformations of material surfaces. The Lagrangian-Averaged Vorticity Deviation (LAVD) \cite{Hall16a} is the state-of-the-art method specifically for elliptic LCS, but is not designed to capture parabolic or hyperbolic LCS. 


\subsection{2D Turbulence}
\begin{figure*}[h!]
\begin{center}
\includegraphics[width=0.929\textwidth,keepaspectratio]{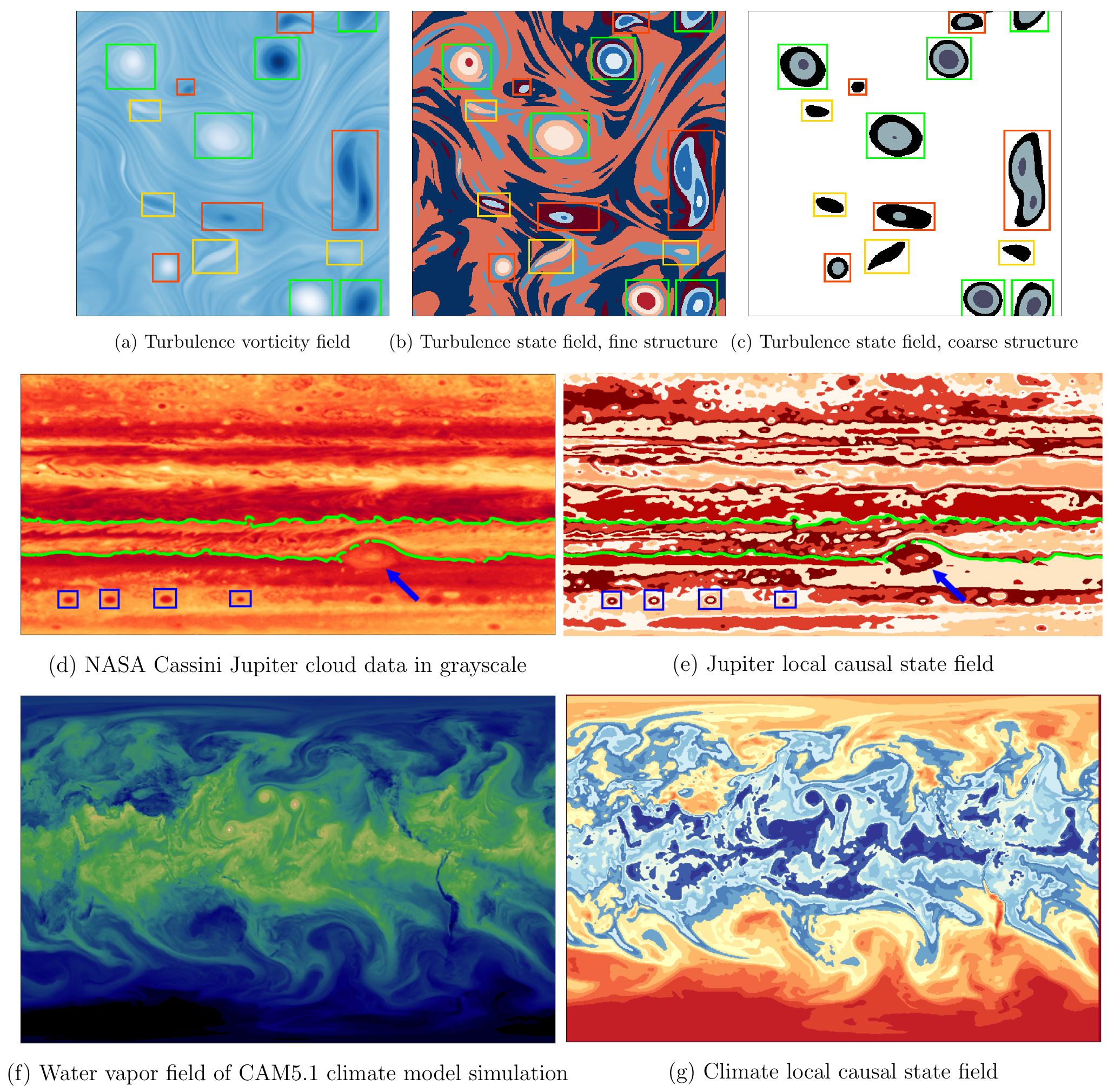}
\caption{
Structural segmentation results for complex fluid flows. 
Image (a) shows the vorticity observable field for the 2D turbulence data set, with (b) and (c) showing the corresponding latent variable local causal state fields. The segmentation in (c) is tuned to isolate vortices against the background potential flow, while the segmentation in (b) captures additional structure of the background. 
Image (d) shows the cloud luminosity observable for the Jupiter data set, with the corresponding local causal state field shown in (e). 
The CAM5.1 water vapor field is shown in (f), with corresponding local causal state field in (g).
Each unique color in the latent variable fields (b), (c), (e), and (g) corresponds to a unique local causal state. 
Full segmentation videos are available on the DisCo YouTube channel \cite{discoyoutube}.
}
\label{fig:science}
\end{center}
\end{figure*}

While still complex and multi-scale, the idealized 2D turbulence data provides the cleanest identification of Lagrangian Coherent Structures using our structural segmentation. Figure~\ref{fig:science} (a) shows a snapshot of the vorticity field, and (b) and (c) show corresponding snapshots from structural segmentations using different reconstruction parameter values. To reveal finer structural details that persist on shorter time scales, Figure~\ref{fig:science} (b) uses $\tau=0.8$ and $K=10$. To isolate the coherent vortices, which persist at longer time scales, Figure~\ref{fig:science} (c) was produced using $\tau=0.0$ and $K=4$. As can be seen in (b), the local causal states distinguish between positive and negative vortices, so for (c) we modded out this symmetry by reconstructing from the absolute value of vorticity. 

All three images are annotated with color-coded bounding boxes outlining elliptic LCS to directly compare with the geodesic and LAVD LCS results from Figure 9, (k) and (l) respectively, in \cite{Hadj17a}. Green boxes are vortices identified by both the geodesic and LAVD methods and red boxes are additional vortices identified by LAVD but not the geodesic. Yellow boxes are new structural signatures of elliptic LCS discovered by the local causal states. 

Because there is a single background state in (c), colored white, all states not colored white can be assigned a semantic label of \texttt{coherent structure} since they satisfy the local causal state definition given in \cite{Rupe18b} as spatially localized, temporally persistent deviations from generalized spacetime symmetries. Significantly, our method has discovered vortices in the observable field (a) as coherent structures due to the shared geometry with the latent field in (c) where they are identified as locally broken symmetries. 


In the finer-scale structural segmentation of (b) we still have states outlining the coherent vortices, as we would expect. If they persist on longer time scales, they will also persist on the short time scale. The additional structure of the background potential flow largely follows the hyperbolic LCS stable-unstable manifolds. Because they act as transport barriers, they partition the flow on either side and these partitions are given by two distinct local causal states with the boundary between them running along the hyperbolic LCS in the unstable direction. For example, the narrow dark blue-colored state in the upper right of (b) indicates a narrow flow channel squeezed between two hyperbolic LCS. 

\subsection{Jupiter}
Figure~\ref{fig:science} (d) shows a snapshot from the Jupiter cloud data, with corresponding structural segmentation snapshot in (e). The Great Red Spot, highlighted with a blue arrow, is the most famous structure in Jupiter's atmosphere. As it is a giant vortex, the Great Red Spot is identified as an elliptic LCS by both the geodesic and LAVD methods \cite{Hadj16a, Hadj17a}. While the local causal states in (e) do not capture the Great Red Spot as cleanly as the vortices in (b) and (c), it does have the same nested state structures as the turbulence vortices.
There are other smaller vortices in Jupiter's atmosphere, most notably the ``string of pearls'' in the Southern Temperate Belt, four of which are highlighted with blue bounding boxes. We can see in (e) that the pearls are nicely captured by the local causal states, similar to the turbulence vortices in (b). 

Perhaps the most distinctive features of Jupiter's atmosphere are the zonal belts. The east-west zonal jet streams that form the boundaries between bands are of particular relevance to Lagrangian Coherent Structure analysis. Figure 11 in \cite{Hadj16a} uses the geodesic method to identify these jet streams as shearless parabolic LCS, indicating they act as transport barriers that separate the zonal belts.
The particular segmentation shown in (e) captures a fair amount of detail inside the bands, but the edges of the bands have neighboring pairs of local causal states with boundaries that extend contiguously in the east-west direction along the parabolic LCS transport barriers. Two such local causal state boundaries are highlighted in green, for comparison with Figure 11 (a) in \cite{Hadj16a}. The topmost green line, in the center of (d) and (e), is the southern equatorial jet, shown in more detail in Figure 11 (b) and Figure 12 of \cite{Hadj16a}. Its north-south meandering is clearly captured by the local causal states.



\subsection{Extreme Weather Events}
The strong qualitative correspondence between local causal state structural segmentation and LCS gives validation that our method can capture meaningful structure in complex spatiotemporal systems. Our aim now is to use the structural segmentation to build extreme weather segmentation masks for climate data. Each event, e.g. hurricanes or atmospheric rivers (ARs), are identified as a unique set of structured behaviors that are captured by the local causal states in a structural segmentation.

For example, a structural segmentation of the water vapor field of the CAM5.1 global atmospheric model is shown on YouTube~\cite{discoyoutube} and in Figure~\ref{fig:science} (e), (f). While signatures of hurricanes and ARs are visually apparent, these events are not uniquely identifiable from the local causal states. 
However, hurricanes and ARs have characteristic structural signatures in other physical fields, including thermodynamic quantities such as temperature and pressure. So it is not surprising that they can not be uniquely identified from a structural segmentation of the water vapor field alone; this would be akin to describing hurricanes as just local concentrations of water vapor. 

In addition to further optimization and scaling, we are currently working on implementing multi-variate local causal state reconstruction to incorporate additional physical fields. 
Using, for example, structural segmentation of vorticity, temperature, pressure, and water vapor fields we will be able to identify hurricanes as high rotation objects with a warm, low pressure core that locally concentrate water vapor. 
Similarly, the inclusion of water vapor transport will help identify ARs, as well as the use of larger lightcone templates that will better capture their large-scale geometry.

Though our structural segmentation requires information from multiple physical observables to identify extreme weather events, the generality of the local causal states will allow us to do this. 
An automated, objective identification of sets of local causal states across the various physical observables that uniquely corresponds to particular extreme weather events will be challenging but, we believe, achievable.  

\section*{Acknowledgements}
Adam Rupe and Jim Crutchfield would like to acknowledge Intel{\small \textregistered} \;for supporting the IPCC at UC Davis. Prabhat and Karthik Kashinath were supported by the Intel\textregistered \;Big Data Center.
This research is based upon work supported by, or in part by, the U. S. Army Research Laboratory and the U. S. Army Research Office under contract W911NF- 13-1-0390, and used resources of the National Energy Research Scientific Computing Center, a DOE Office of Science User Facility supported by the Office of Science of the U.S. Department of Energy under Contract No. DE-AC02-05CH11231.

\bibliographystyle{ieeetr}
\bibliography{tusewe}

\end{document}